%% file: main.tex
\documentclass[twocolumn]{aastex63}
\usepackage{longtable}
\usepackage{multirow}

\makeatletter
\newcommand*{\bigcdot}{}% Check if undefined
\DeclareRobustCommand*{\bigcdot}{%
  \mathbin{\mathpalette\bigcdot@{}}%
}
\newcommand*{\bigcdot@scalefactor}{.5}
\newcommand*{\bigcdot@widthfactor}{1.15}
\newcommand*{\bigcdot@}[2]{%
  \sbox0{$#1\vcenter{}$}% math axis
  \sbox2{$#1\cdot\m@th$}%
  \hbox to \bigcdot@widthfactor\wd2{%
    \hfil
    \raise\ht0\hbox{%
      \scalebox{\bigcdot@scalefactor}{%
        \lower\ht0\hbox{$#1\bullet\m@th$}%
      }%
    }%
    \hfil
  }%
}
\makeatother

\shorttitle{Hot Stars With Kepler Planets}
\shortauthors{Louden et al.}
\graphicspath{{./}{figures/}}

\begin{document}

\title{Hot Stars With Kepler Planets Have High Obliquities\footnote{The data presented herein were obtained at the W. M. Keck Observatory, which is operated as a scientific partnership among the California Institute of Technology, the University of California and the National Aeronautics and Space Administration. The Observatory was made possible by the generous financial support of the W.M.~Keck Foundation.}}

\correspondingauthor{Joshua N.\ Winn}
\email{jnwinn@princeton.edu}

\author[0000-0003-3179-5320]{Emma M.\ Louden}
\affiliation{Department of Astrophysical Sciences, Princeton University, 4 Ivy Lane, Princeton, NJ 08544, USA}

\author[0000-0002-4265-047X]{Joshua N.\ Winn}
\affiliation{Department of Astrophysical Sciences, Princeton University, 4 Ivy Lane, Princeton, NJ 08544, USA}

\author[0000-0003-0967-2893]{Erik A.\ Petigura}
\affiliation{Department of Physics \& Astronomy, University of California Los Angeles, Los Angeles, CA 90095, USA}

\author[0000-0002-0531-1073]{Howard Isaacson}
\affiliation{501 Campbell Hall, University of California at Berkeley, Berkeley, CA 94720, USA}

\author[0000-0001-8638-0320]{Andrew W.\ Howard}
\affiliation{Department of Astronomy, California Institute of Technology, Pasadena, CA, USA}

\author[0000-0003-1298-9699]{Kento Masuda}
\affiliation{Department of Earth and Space Science, Osaka University, Osaka 560-0043, Japan} 

\author[0000-0003-1762-8235]{Simon Albrecht}
\affil{Stellar Astrophysics Centre, Department of Physics and Astronomy, Aarhus University, Ny Munkegade 120, 8000 Aarhus C, Denmark}

\author[0000-0002-6115-4359]{Molly R.\ Kosiarek}
\altaffiliation{NSF Graduate Student Research Fellowship}
\affiliation{Department of Astronomy and Astrophysics, University of California, Santa Cruz, CA 95064, USA}

\begin{abstract}

It has been known for a decade that hot stars with hot Jupiters tend
to have high obliquities. Less is known about the degree of spin-orbit 
alignment for hot stars with other kinds of planets.
Here, we re-assess
the obliquities of hot {\it Kepler} stars with transiting planets smaller
than Neptune, based on spectroscopic measurements of their projected
rotation velocities ($v\sin i$).
The basis of the method is that a lower obliquity --- all other things being
equal --- causes $\sin i$ to be closer to unity and increases the value of $v\sin i$.
We sought evidence for this effect using
a sample of 150 {\it Kepler} stars with effective temperatures between
5950 and $6550$\,K and a control sample
of 101 stars with matching spectroscopic properties
and random orientations.
The planet hosts have systematically
higher values of $v\sin i$ than the control stars, but not by enough
to be compatible with perfect spin-orbit alignment.
The mean value of $\sin i$ is $0.856\pm 0.036$,
which is 4-$\sigma$ away from unity (perfect alignment),
and 2-$\sigma$ away from $\pi/4$ (random orientations).
There is also evidence that the hottest stars have a broader obliquity distribution:
when modeled separately,
the stars cooler than $6250$\,K have $\langle \sin i \rangle = 0.928 \pm 0.042$
while the hotter stars are consistent with random orientations.
This is similar to the pattern previously
noted for stars with hot Jupiters.
Based on these results, obliquity excitation for early-G and late-F stars 
appears to be a general outcome of star and planet formation, rather than
being exclusively linked to hot Jupiter formation.

\end{abstract}

\section{Introduction} \label{sec:intro}

Sometimes, the equatorial
plane of a star is grossly misaligned with the orbital
plane of at least one of its planets
\citep[see, e.g.,][]{WinnFabrycky2015,Triaud2018}.
The reasons for these high stellar obliquities are unclear.
Most of the available data are for stars with hot
Jupiters. Early on, it became clear that cool stars ($T_{\rm eff} \lesssim 6000$\,K) with hot Jupiters
tend to have low obliquities \citep{FabryckyWinn2009}.
Among those stars, observations of high obliquities have mainly
been restricted to those with
wider-orbiting giant planets ($a/R_\star \gtrsim 8$).
In contrast, hotter stars with hot Jupiters
have a much broader obliquity distribution \citep{Schlaufman2010,Winn2010}.

Many of the theories that have been offered to
explain these results invoke formation pathways for hot Jupiters in which the
planet's orbital plane is tilted away from
the protoplanetary disk plane.
Good alignment might eventually be restored by tidal dissipation,
but only for 
cool stars with especially close-orbiting giant
planets, owing to stronger tidal dissipation
and more rapid magnetic braking \citep{Winn2010,Dawson2014}.
It remains possible, though, that high obliquities are
unrelated to hot Jupiter formation and instead reflect
more general processes in star and planet
formation. One way to make progress is to
measure the obliquities of stars with different types
of planets, including smaller and wider-orbiting planets
than hot Jupiters.

The {\it Kepler} survey provided a sample of about 4{,}000
transiting planets around FGKM host stars, the majority of which have
sizes between 1 and 4\,$R_\oplus$ and orbital
periods ranging from 1 to 100 days \citep{Borucki2017,Thompson2018}.
In most cases, measuring the obliquities
of individual stars via the Rossiter-McLaughlin effect
is impractical, owing to the faintness of the star
and the small size of the planet.  But the sample is large
enough for statistical probes of the obliquity distribution.
In particular, since the planetary orbits are all being viewed at high
inclination with respect to the line of sight (a requirement
for transits to occur), any constraints on the inclination
distribution of the stellar rotation axes are also
constraints on the stellar obliquity distribution.

\citet{Mazeh+2015} performed one such study, based on measurements
of rotationally-induced photometric variability. They found
clear evidence
that stars cooler than about 6000\,K have low obliquities,
as well as suggestive evidence that hotter stars
have a broad range of obliquities, with caveats to be
discussed later in this paper.

\cite{Winn+2017} and \cite{MunozPerets2018}
performed studies using measurements of the projected
rotation velocities ($v\sin i$).
Both sets of authors examined the cases in which measurements
of $v\sin i$, rotation period, and stellar radius are available,
to obtain constraints on $\sin i$.  The results were generally
consistent with low obliquities $(\lesssim\,30^\circ)$.
A limitation of these studies was that the sample of stars with
detected rotation periods
may suffer from biases
that favor low-mass stars, high inclinations, and relatively short rotation periods,
all of which facilitate the detection of a photometric rotation signal.

\cite{Winn+2017} also compared the $v\sin i$ distributions
of planet hosts and samples of stars chosen without
regard to planets. The planet hosts had systematically higher
values of $v\sin i$, again suggesting low obliquities. 
However, the comparison stars were drawn from heterogeneous sources,
some of which may have been biased against high-inclination stars and
rapid rotators.
This called into question the key assumptions that the comparison
stars are randomly oriented and have the
same distribution of rotation velocities
as the planet hosts.
The work presented here is a new application of this method
with an improved control sample.

The rest of this paper is organized as follows.
Section 2 describes our observations of the candidate control stars.
Section 3 compares
the spectroscopic properties of the planet hosts and the control
stars.  Section 4 presents two statistical tests for
differences between the $v\sin i$ distributions of the two
samples.  Section 5 describes a simple model that was used to
characterize the obliquity distribution of the planet hosts.
Section 6 summarizes and describes possible implications
for theories of obliquity or inclination excitation.

\section{Observations} \label{sec:obs}

The best stars for this type of study are
early-G and late-F main-sequence stars.
Cooler stars typically rotate too slowly to permit reliable measurements
of $v\sin i$, and hotter stars are not well represented in the {\it Kepler}
sample of planet-hosting stars.
We drew the data for the planet hosts from
the California-{\it Kepler} Survey \citep[CKS,][]{Petigura+2017,Johnson2017}.
The CKS team performed Keck/HIRES spectroscopy
of 1{,}305 stars with transiting planets, of which several hundred have
spectral types in the desired range.
They provided precise determinations
of the effective temperature ($T_{\rm eff}$), surface gravity ($\log g$),
iron metallicity ([Fe/H]), and projected rotation
velocity ($v\sin i$).

We needed to construct a control sample 
as similar as possible to the {\it Kepler} planet hosts,
but selected without regard to rotation rate
or orientation.
Only with such a sample can any systematic differences
in $v\sin i$ between the planet hosts
and control stars be attributed to the
obliquity distribution of the planet hosts.
We also wanted to observe the control stars
with the same instrument as the planet hosts,
and use the same software to analyze the spectra.
This is important because
measurements of $v\sin i$ are subject to systematic errors
related to instrumental resolution and treatment of other line-broadening
mechanisms.

We selected candidate control stars
based on low-resolution spectroscopy of the {\it Kepler} field
by the LAMOST team \citep{Ren+2016}.
We defined a similarity metric between two stars:
\begin{equation}
  D^2 = \left( \frac{\Delta T_{\rm eff}}{100\,{\rm K}} \right)^2 +
  \left( \frac{\Delta\log g}{0.10\,{\rm dex}} \right)^2 +
    \left( \frac{\Delta{\rm [Fe/H]}}{0.10\,{\rm dex}} \right)^2.
\end{equation}
The quantities in the denominators are typical
LAMOST uncertainties. We chose a trial value
of $m_{\rm lim}$, the limiting apparent magnitude in the
{\it Kepler} bandpass. For each
CKS star in the desired range of effective temperatures, we
selected the LAMOST star with the minimum $D$ and $m<m_{\rm lim}$.
Then, we adjusted $m_{\rm lim}$ to be the
brightest possible value for which two-sided Kolmogorov-Smirnov tests
did not reject the hypotheses that the distributions
of $T_{\rm eff}$, $\log g$ and [Fe/H] are the same for the CKS
stars and the candidate control stars.
This turned out to be $m_{\rm lim} = 11.1$, approximately
3 magnitudes brighter than the limiting magnitude
of the planet hosts.

We observed 188 candidate control stars with Keck/HIRES
during the summer of 2018. The observations were spread
out over several days, amounting to a total of about one half-night
of Keck time. We used the same instrumental setup, observing
protocols, data reduction software, and analysis procedures
that were used by the CKS.
In particular, the basic spectroscopic parameters of each star
were determined with SpecMatch \citep{Petigura2015},
for which the CKS team 
demonstrated an internal precision of 60~K in $T_{\rm eff}$,
0.10~dex in $\log g$, 0.04~dex in [Fe/H], and 1.0 km\,s$^{-1}$ in $v\sin i$. 
The latest version of SpecMatch was applied to the Keck/HIRES spectra of
both the planet hosts and the control stars as a single batch job,
to ensure homogeneity in the analysis method.\footnote{In Paper~I
of the CKS series of publications
\citep{Petigura+2017}, the tabulated spectroscopic
parameters are based on an average of the results 
obtained with two different analysis codes: SpecMatch, and
{\tt SME@XSEDE}. For our study, we used only SpecMatch.}

Nineteen of the candidate control stars turned out to be spectroscopic binaries
and were discarded from the sample.
Following the same quality control procedures as the recent CKS
study by \citet{FultonPetigura2018},
we also eliminated from consideration
any star for which
the \citet{Gaia2018} geometric parallax is not available,
has a precision lower than 10\%,
or disagrees with the spectroscopic parallax by more than 4-$\sigma$.

\section{Sample construction} \label{sec:samples}

Because the selection of candidate control stars was based on low-resolution
data, in many cases the stars turned out to
have spectroscopic parameters far away from those
of the planet hosts. To construct samples
with overlapping properties,
we restricted both the planet hosts and the control stars to have
SpecMatch parameters satisfying
\begin{eqnarray*}
5950\,{\rm K} < &T_{\rm eff}& < 6550\,{\rm K},\\
3.95 < &\log g& < 4.45,~{\rm and}\\
-0.3 < &{\rm [Fe/H]}& < 0.3.
\end{eqnarray*}
Since we were interested in the obliquities of stars with small
planets --- and not hot Jupiters --- we only included stars
having at least one planet smaller than 4\,$R_\oplus$.
This led to our final samples of 150 planet hosts and 101 control stars.
The SpecMatch parameters of these stars are given in Tables~2 and 3,
which appear at the end of the paper.
Figure~1 %\ref{fig:RadiusPeriod}
shows the radius/period distribution of the known
transiting planets for all the planet hosts in our sample.

\begin{figure}
\setcounter{figure}{0}
\label{fig:RadiusPeriod}
\begin{center}
\includegraphics[width=0.45\textwidth]{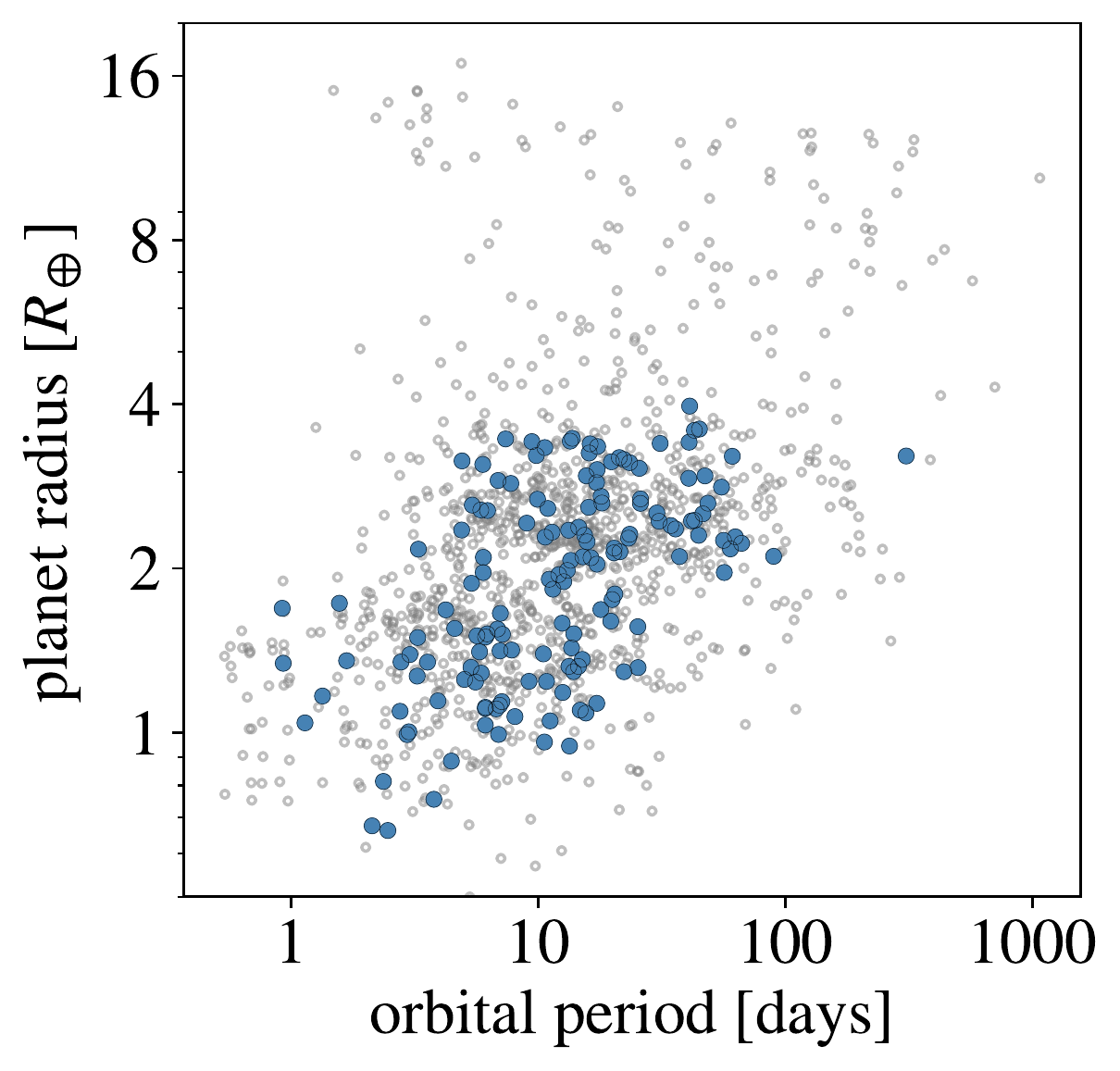}
\end{center}
\caption{Planetary radius and orbital period, for all the
known transiting planets associated with the 153 planet
hosts in our sample.}
\end{figure}

\begin{figure*}
\label{fig:SampleComparison}
\begin{center}
\includegraphics[width=0.9\textwidth]{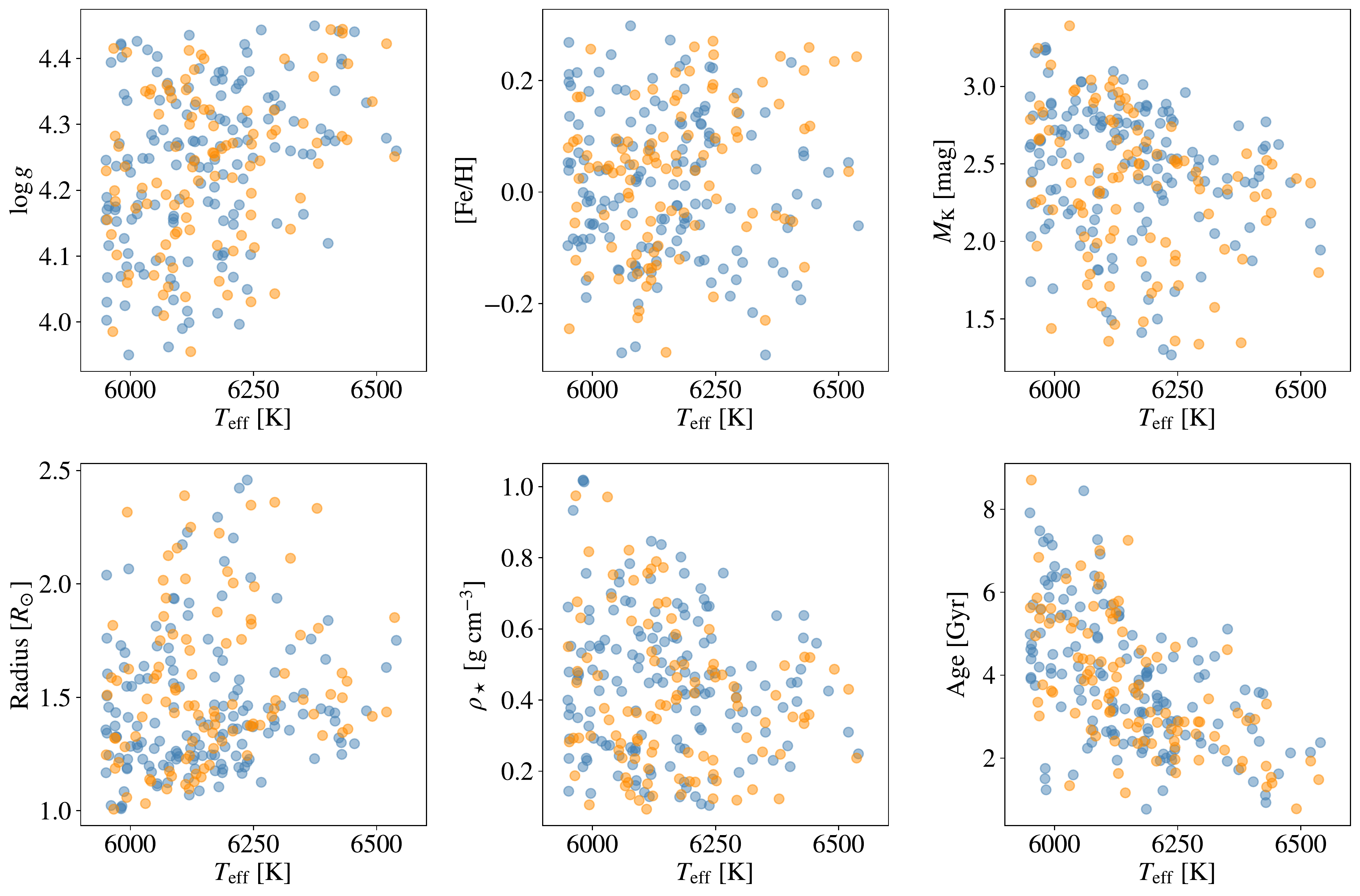}
\end{center}
\caption{Comparison between the properties of
the planet hosts (blue) and control stars (red).
The precision of the measurements of $T_{\rm eff}$,
$\log g$, [Fe/H], and $v\sin i$ is 
60~K, 0.10~dex, 0.04~dex, and 1.0 km\,s$^{-1}$, respectively.
In the the upper right panel, the vertical coordinate
is the absolute magnitude based on the 2MASS
apparent magnitude $m_K$ and the Gaia DR2 parallax, with no allowance
for extinction.
The radius, mean density, and age are
from fits to MIST isochrones,
and have typical internal uncertainties
of 3\%, 6\%, and 30\%, respectively.}
\end{figure*}

We used two-sided Kolmogorov-Smirnov tests
to check on the ``null hypothesis'' that
the planet hosts and control stars have
spectroscopic parameters drawn from the same
parent distribution.
The null hypothesis cannot be ruled out for
$T_{\rm eff}$ ($p=0.57$),
$\log g$ ($p=0.99$), or
[Fe/H] ($p=0.44$).
Likewise, the Anderson-Darling test
cannot reject the null hypothesis
for any of those three parameters ($p=0.66$, 0.99,
and 0.43, respectively).\footnote{For these tests,
as well as the other nonparametric tests described
in this paper, the $p$-values were determined by bootstrap resampling,
not by analytic approximations.}

The preceding tests did not find any differences in the
distributions of individual parameters, but are not capable
of checking for differences in the joint distribution
of two parameters.  For this,
we performed the two-dimensional generalization of the
Kolmogorov-Smirnov test 
described by \cite{PressTeukolsky1988}, which they attributed
to earlier work by \citet{FasanoFranceschini1987} and \citet{Peacock1983}.
We tested the joint distributions
of $(T_{\rm eff}, \log g)$,
$(T_{\rm eff}, {\rm [Fe/H]})$, and $(\log g, {\rm [Fe/H]})$.
In all three cases, the test result was compatible with the hypothesis
that the parameters are drawn from the same joint distribution ($p>0.3$).
While these tests are only 2-d and not 3-d, and share the same
shortcomings as the original KS test \citep[see, e.g.,][]{FeigelsonBabu2012},
they give us some confidence that the control stars are similar
to the planet hosts.

Figure~2 % \ref{fig:SampleComparison}
shows the distributions of the spectroscopic parameters
and other parameters of interest.
This includes the $K$-band absolute magnitude, computed
from the 2MASS apparent magnitude \citep{Cutri+2003}
and the Gaia
parallax \citep{Gaia2018} without
any correction for extinction.
The other parameters depicted are
the mass, radius, mean density, and age
of the stars, based on fitting the spectroscopic
parameters to the MIST stellar-evolutionary models \citep{Choi2016},
using the method described by \cite{FultonPetigura2018}.

\section{Model-Independent Tests} \label{sec:model-independent}

\begin{figure*}
\label{fig:VsiniTeff}
\begin{center}
\includegraphics[width=0.8\textwidth]{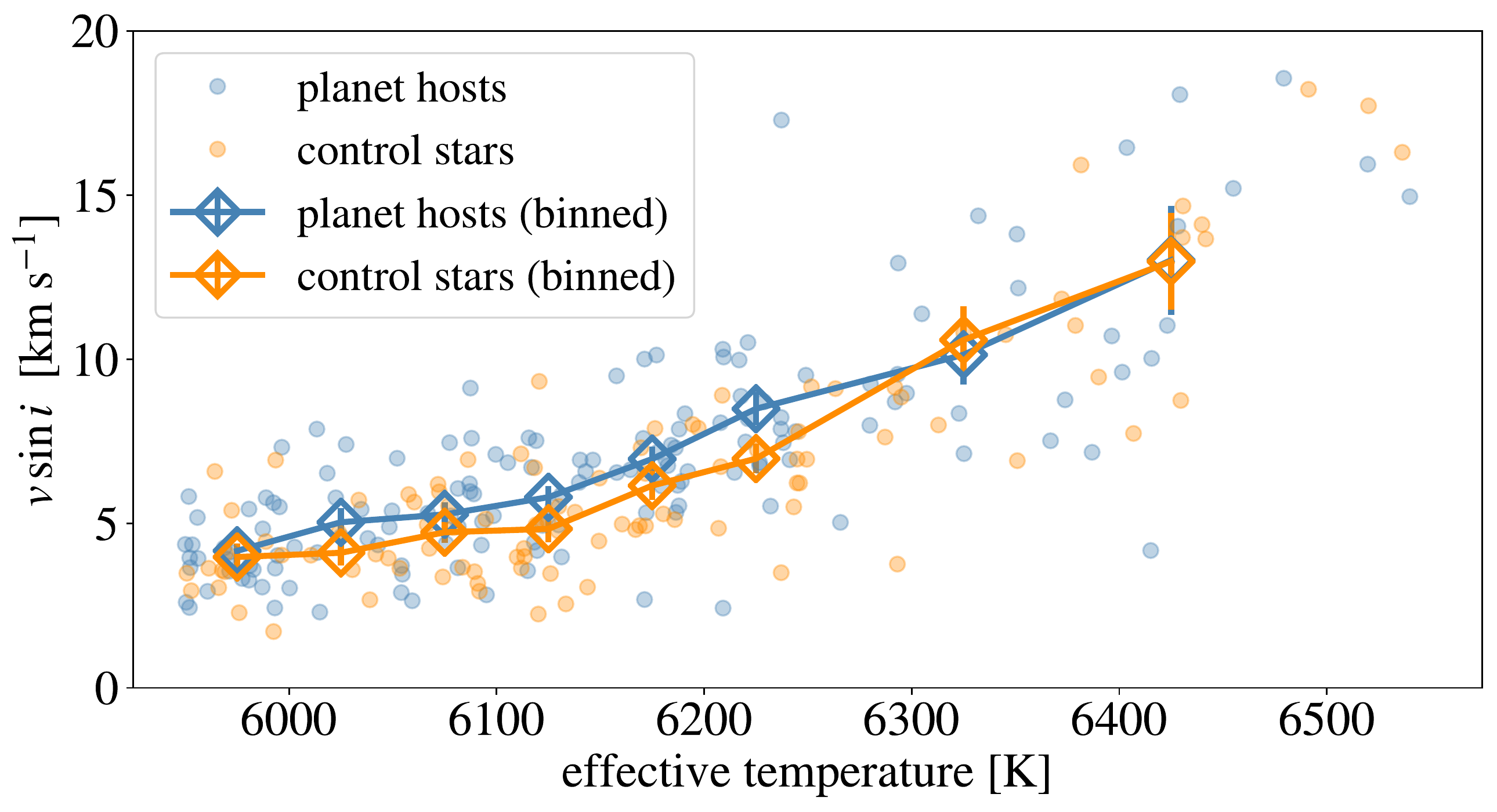}
\end{center}
\caption{Measurements of projected rotation velocity versus effective temperature,
for planet hosts (blue) and control stars (orange).
The diamonds are binned values of $v\sin i$.
For effective temperatures cooler than 6250\,K, the planet hosts
have higher mean values of $v\sin i$ than the control stars, indicating
a tendency toward spin-orbit alignment.}
\end{figure*}

Figure~3
% \ref{fig:VsiniTeff}
shows the projected rotation velocity as a
function of effective temperature, both for individual stars and
for averages within temperature bins.
The bins were chosen to have a width of 50\,K for
stars cooler than 6250\,K, and 100\,K for the less
numerous hotter stars.
For both samples, the average value of
$v\sin i$ rises with $T_{\rm eff}$, as expected;
this temperature range spans the
well-known ``Kraft break'' above which stars are observed
to rotate faster \citep{Struve1930,Kraft1967}.  This
trend is attributed to the reduced rate of magnetic braking
for hot stars that lack thick outer convective envelopes.

It appears from Figure~3 that the relatively cool planet-hosting
stars ($T_{\rm eff} < 6250$\,K) tend to 
have higher $v\sin i$ values than
the control stars.  This is a sign that these planet hosts have systematically
higher values of $\sin i$ and therefore have low obliquities.
We performed two statistical tests to quantify the difference in the $v\sin i$
distributions.

First, we performed the two-dimensional Kolmogorov-Smirnov
test referenced earlier, using $T_{\rm eff}$ and $v\sin i$ as the
two dimensions.  The null hypothesis that the planet hosts
and control stars have values of these two parameters drawn
from the same joint distribution is assigned $p=0.028$.  When applied only to the
planet hosts and control stars with $T_{\rm eff} < 6250$\,K,
the same test gives $p=0.0034$, representing a stronger rejection
of the null hypothesis.

The second test was based on the observation that
the planet hosts have a mean $v\sin i$
that exceeds that of the control stars in all of the first
6 temperature bins shown in Figure~3.
How often would differences at this level
occur by chance, if $T_{\rm eff}$ and
$v\sin i$ for all the stars were drawn from the same two-dimensional distribution?
We answered this question through a Monte Carlo procedure.
We quantified the difference between the two distribution with the statistic
\begin{equation}
   S \equiv
    \sum_{n=1}^8
    \frac{
      \langle v\sin i\rangle_{{\rm p},n} - \langle v\sin i\rangle_{{\rm c},n}
    }{
      \sqrt{ \sigma^2_{{\rm p},n} + \sigma^2_{{\rm c},n} }
    },
\end{equation}
where $\langle v\sin i\rangle_n$ is the mean value $v\sin i$ within the $n$th temperature
bin; $\sigma_n$ is the corresponding standard deviation of the mean; and ``p'' and
''c'' refer to the planet sample and the control sample, respectively.
The real data have $S_{\rm obs} = 8.3$.
To create simulated data sets, we combined the 150 planet hosts and 101 control
stars to form a combined sample of 251 stars, and then randomly drew (with replacement)
150 members of the combined sample to serve as ``planet hosts'' and 101 members
to serve as ``control stars.''  By construction, the simulated data sets
have parameters that are drawn from the same joint distribution.  We computed
the $S$ statistic for each of $10^5$
simulated data sets; in no case did we find $S>S_{\rm obs}$.  Therefore, according
to this test, $p<10^5$.

These model-independent tests confirmed the visual impression
that the $v\sin i$ distributions of the planet hosts and control stars
are significantly different, at least for the stars with $T_{\rm eff}<6250$\,K.
In the following sections, we use a simple model
to quantify the resulting
constraints on the obliquity distribution of the planet-hosting stars.

\section{A Simple Model} \label{sec:analysis}

\begin{figure*}
\label{fig:VsiniTeff_withModels}
\begin{center}
\includegraphics[width=0.8\textwidth]{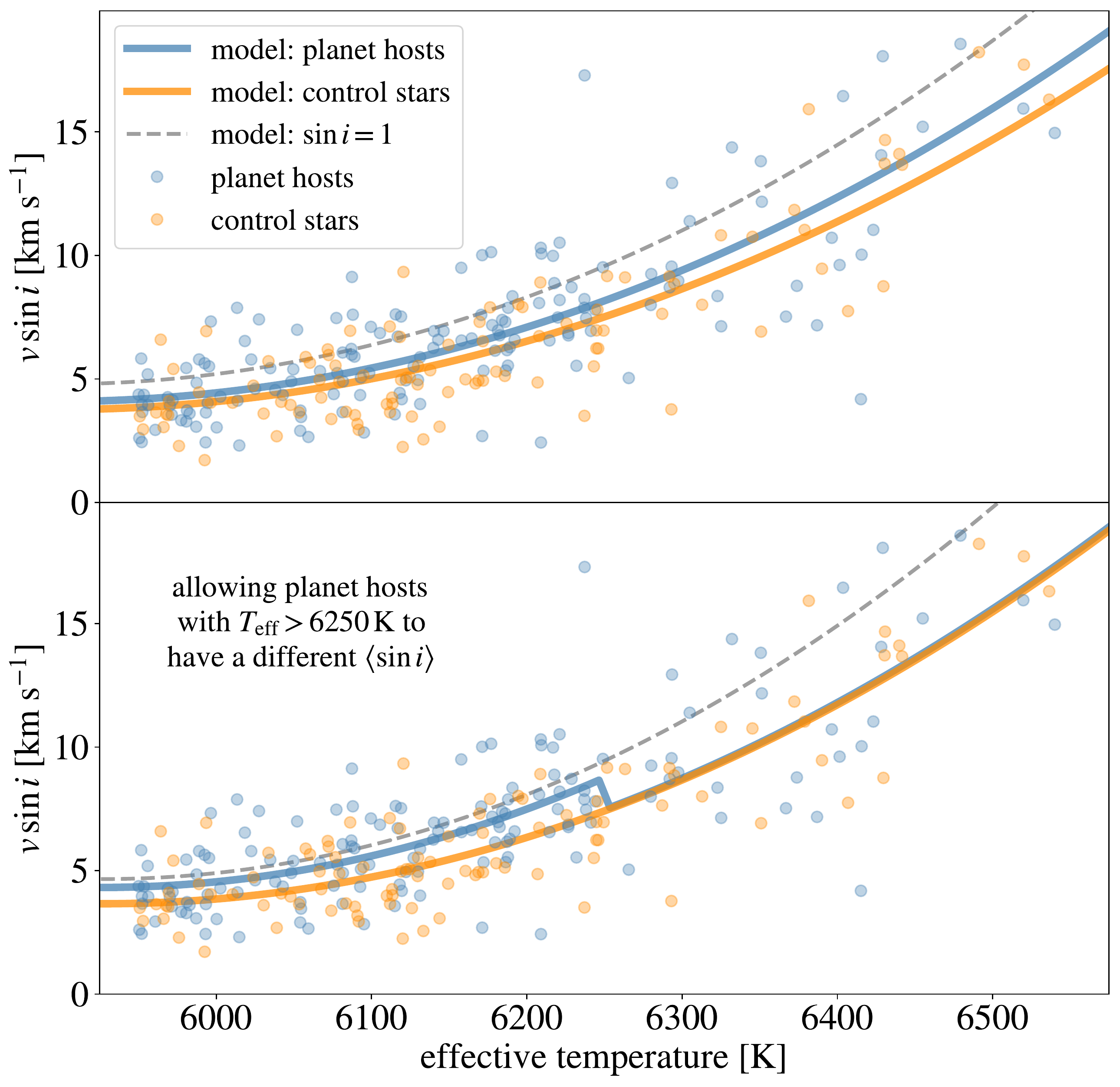}
\end{center}
\caption{Measurements of projected rotation velocity versus effective temperature,
for planet host (blue) and control stars (orange).
The curves illustrate the best-fitting models. In the top panel,
all the planet hosts are assumed to have the same value of $\langle \sin i\rangle$.
In the bottom panel, the hosts cooler than 6250\,K were
allowed to have a different value of $\langle \sin i \rangle$
from the hosts hotter than 6250\,K.
In both panels, the gray dashed curve is the
mean rotation velocity $\langle v\rangle$,
the blue curve is $\langle v\rangle \langle \sin i\rangle$
fitted to the planet hosts,
and the orange curve is $\langle v\rangle\times \pi/4$ fitted
to the control stars.}
\end{figure*}

\subsection{Premises}

Our model is based on the following premises:
\begin{enumerate}
    \item A star's rotation velocity $v$ and inclination $i$ are
    independent variables.  This seems uncontroversial, since the
    rotation velocity is an intrinsic quantity,
    while the inclination depends on our arbitrary position within the galaxy.

    \item For any value of the effective temperature,
    the control stars and the planet hosts have the
    same distribution of rotation velocities.  This is justified
    by the sample construction and comparisons presented in Section~3.

    \item The mean rotation velocity $\langle v\rangle$
    is a quadratic function of effective temperature. This
    is a simplifying assumption based on the trend observed in Figure~3.

    \item The measurements of $v\sin i$ for
    the control stars and the planet hosts
    are subject to the same systematic uncertainties.  Ensuring this is the case
    was the motivation for obtaining all the spectra with the same instrument and analyzing them with
    the same code.

    \item The control stars are randomly oriented in space.
\end{enumerate}
To these, we add a sixth premise, and consider two different cases:
\begin{itemize}
    \item[6a.] The obliquities of the transiting planet hosts
    are all drawn from the same distribution.
\setcounter{enumi}{4}
    \item[6b.] There are two different obliquity distributions:
    one for hosts cooler than 6250\,K,
    and one for hosts hotter than 6250\,K.
\end{itemize}
The second case is inspired by the appearance of Figure~3,
as well as the fact that the obliquity distribution of hot Jupiter hosts
has been observed to broaden as the temperature is increased past 6250\,K, the
approximate location of the Kraft break.

The only aspect of the obliquity distribution that is well
constrained by the data is $\langle \sin i\rangle$, the mean
value of $\sin i$ for the planet hosts.
For this reason, our models include $\langle \sin i\rangle$ as
a free parameter but do not 
adopt a particular functional form for the obliquity distribution.
A population of randomly oriented stars would have
$\langle \sin i\rangle = \pi/4 \approx 0.785$, and
a population of transiting-planet hosts with low obliquities
would have $\langle \sin i\rangle \approx 1$.

We fitted a single model to all of the stars, both the planet
hosts and the control stars.  For all the stars,
the mean rotation velocity in the model is
    \begin{equation}
     \langle v\rangle(\tau) = c_0 + c_1 \tau + c_2 \tau^2,
    \end{equation}
where
\begin{equation}
     \tau \equiv \frac{T_{\rm eff} - 6250\,{\rm K}}{300\,{\rm K}}
\end{equation}
varies from $-1$ to $+1$, and $c_0$, $c_1$, and $c_2$ are free parameters.
The mean $v\sin i$ value in the model depends
on whether the star is a control star or a planet host:
\begin{eqnarray}
\langle v\sin i\rangle_n &=& \langle v\rangle_n \times \frac{\pi}{4}~~({\rm control~stars}) \\
\langle v\sin i\rangle_n &=& \langle v\rangle_n \times \langle \sin i\rangle~~({\rm planet~hosts}),
\end{eqnarray}
where we have used the fact that $v$ and $\sin i$ are
uncorrelated.
Thus, in this model, the polynomial coefficients are constrained
by all of the stars, and the $\langle \sin i\rangle$ parameter
is constrained by the planet hosts.

The goodness-of-fit statistic was taken to be
\begin{equation}
    \chi^2 = \sum_{n=1}^{251}
    \left(
    \frac{ v\sin i_{{\rm obs}, n} - \langle v\sin i\rangle_{{\rm calc}, n} }
         {1\,{\rm km\,s}^{-1}}
    \right)^2,
\end{equation}
where $v\sin i_{{\rm obs}, j}$ is the observed value of $v\sin i$ of the $n$th star,
$\langle v\sin i\rangle_{{\rm calc}, i}$ is the mean value of $v\sin i$ calculated
according to the model,
and $1\,{\rm km\,s}^{-1}$ is the measurement uncertainty.

\begin{figure}
\label{fig:Posterior}
\begin{center}
\includegraphics[width=0.45\textwidth]{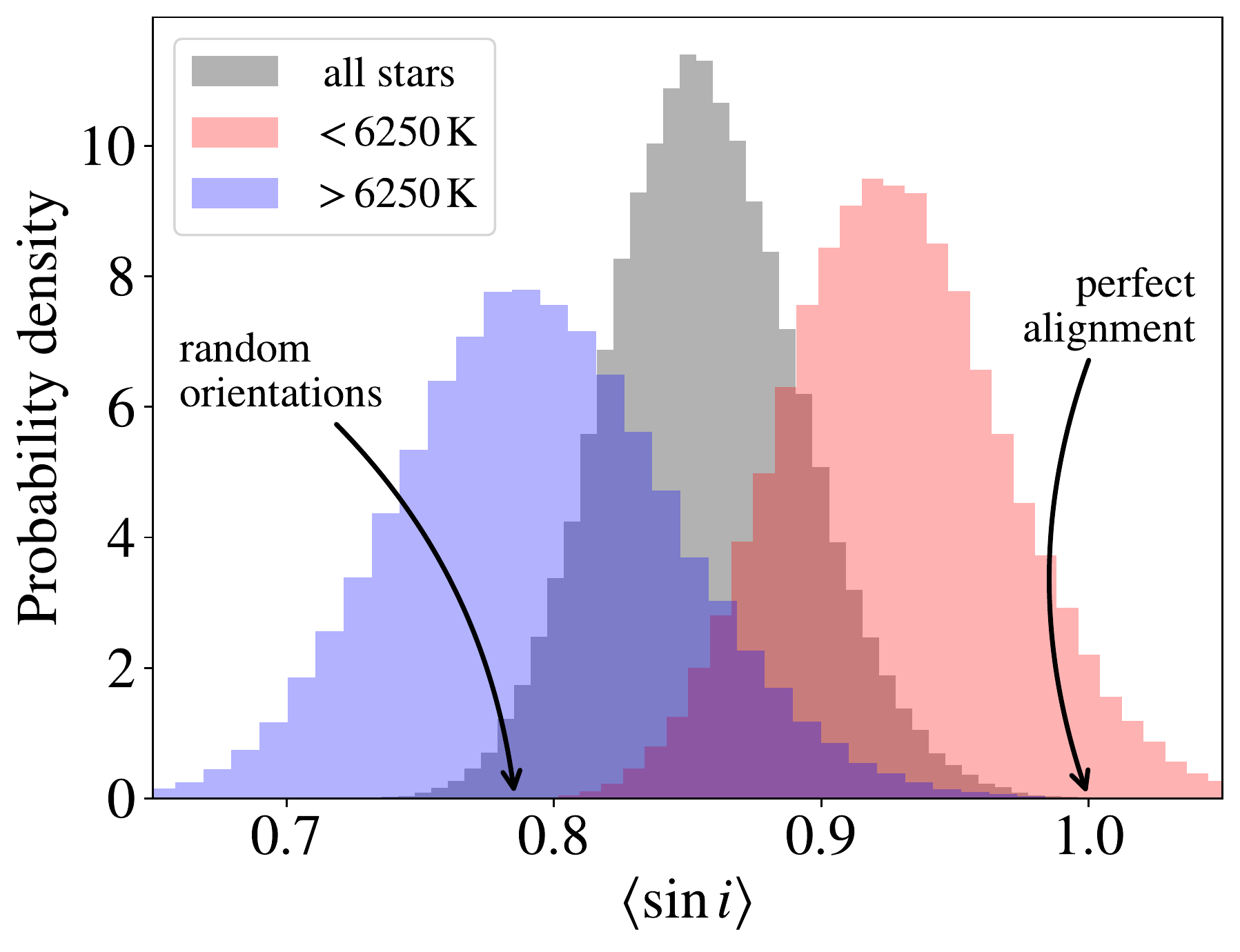}
\end{center}
\caption{Posterior probability distributions for $\langle \sin i\rangle$,
marginalized over all other parameter values. The gray curve shows the
case in which the obliquities of all the planet hosts were assumed to be drawn from the same
distribution. The red and blue curves shows the case in which the stars
with $T_{\rm eff}<6250$\,K were allowed to have a different obliquity
distribution from the stars with $T_{\rm eff}>6250$\,K.}
\end{figure}

\subsection{Results}

For the case of a single obliquity distribution (premise 5a),
the best-fitting model has $\langle \sin i \rangle = 0.856$
and $\chi^2_{\rm min}= 935$, with 247 degrees of freedom
(251 data points and 4 free parameters).  The model does not
fit the data points to within the measurement uncertainties,
nor should we expect it to fit so well.  An individual measurement of $v\sin i$
departs from the calculated $\langle v\sin i\rangle$
not only because of the measurement uncertainty,
but also because of the intrinsic dispersion in the rotation velocities and
the dispersion in $\sin i$.  These deviations
are drawn from different distributions, neither of which is known well.
For this reason, we used a bootstrap procedure to establish the
confidence intervals for the model parameters.

We created $10^5$ simulated data sets, each with
the same number of planet hosts and control
stars as the real data set, by drawing
data points randomly (with repetitions allowed) from
the real data.  The model was fitted to each
simulated data set by minimizing the $\chi^2$ statistic.
The resulting collection
of $10^5$ parameter sets was interpreted as 
a sampling from the joint probability density of
the parameter values.

For the case of a single obliquity distribution (premise 6a),
the bootstrap procedure gave
$\langle \sin i\rangle = 0.856\pm 0.036$, 
where the uncertainty interval encompasses
68\% of the bootstrap simulation results.
For the case of two different obliquity distributions (premise 6b),
the stars cooler than 6250\,K have
$\langle \sin i\rangle = 0.928\pm 0.042$.
The higher value obtained in this case
implies a stronger tendency toward spin-orbit alignment;
indeed, the result differs by only 1.7-$\sigma$ from
the condition of perfect alignment.
Conversely, the stars hotter than 6250\,K have
$\langle \sin i\rangle = 0.794\pm 0.052$,
which is consistent with random orientations ($\pi/4\approx 0.785$).
Table 1 gives the results for all the parameters.
Figure~4 show the best-fitting model curves,
and Figure~5 shows the probability distributions for the key
parameters.

\begin{deluxetable}{ccc}
\label{tbl:allparameters}
\tablecaption{Parameter values.}
\tablehead{
  Parameter & Single obliquity & Two obliquity \\
   value    & distribution     & distributions
}
\startdata
\multirow{2}{*}{$\langle \sin i\rangle$} & \multirow{2}{*}{$0.856\pm 0.036$ } & 
  $0.928\pm 0.042$, $<$6250\,K \\
  & & $0.794\pm 0.052$, $>$6250\,K  \\
$c_0$ & $9.57\pm 0.29$  & $9.44\pm 0.28$ \\
$c_1$ & $8.01\pm 0.54$  & $8.87\pm 0.61$ \\
$c_2$ & $3.30\pm 0.62$  & $4.05\pm 0.62$
\enddata
\end{deluxetable}

\subsection{von-Mises Fisher distribution}

\begin{figure}
\label{fig:vmfdist}
\begin{center}
\includegraphics[width=0.5\textwidth]{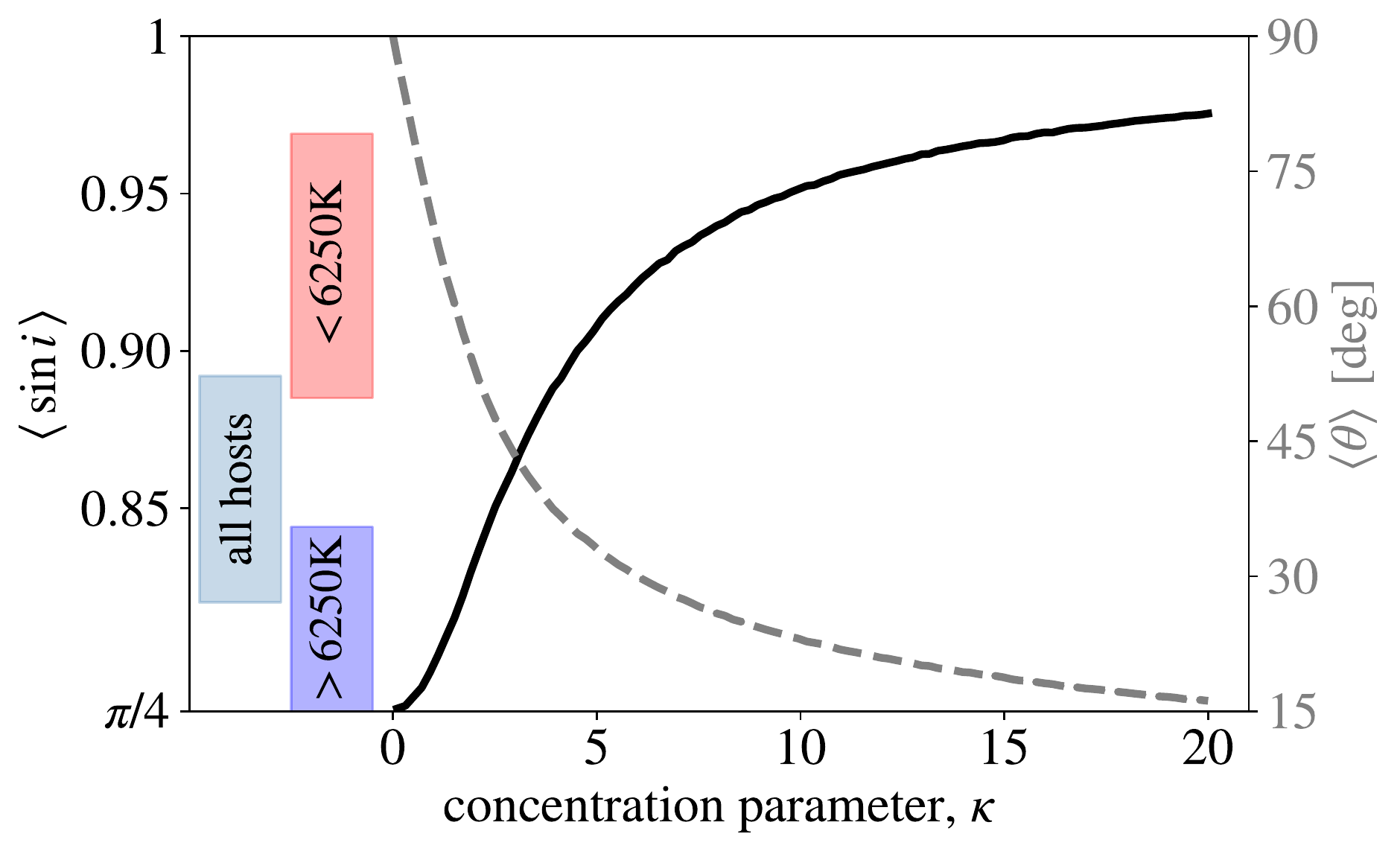}
\end{center}
\caption{Relationship between the concentration parameter $\kappa$ of the von-Mises Fisher
distribution and the mean values of  $\sin i$ (solid black line, left-side axis)
and obliquity (gray dashed line, right-side axis). On the left,
the colored bars indicate the 1-$\sigma$ allowed
ranges of $\langle\sin i\rangle$ for the planet hosts, using the model
described in Section 5.}
\end{figure}

Further steps are needed to obtain quantitative constraints on the obliquity distribution,
because $\sin i$ is only one aspect of the obliquity.  The other aspect is
the position angle $\Omega$ of the projection of the spin axis onto the orbital plane. The
relationship is
\begin{equation}
    \sin i = \sqrt{1 - \sin^2\theta \cos^2\Omega}.
\end{equation}
Even though our model is not committed to a specific shape
for the obliquity distribution, we find it useful to interpret the results with reference to a
von-Mises Fisher (vMF) distribution,
\begin{equation}
    p(\hat{n}_\star) \propto
    \exp(\kappa\hat{n}_\star\!\bigcdot\!\,\hat{n}_{\rm o}),
\end{equation}
where $\hat{n}_\star$ and $\hat{n}_{\rm o}$ are the
unit vectors in the directions of the spin axis
and the orbital axis, respectively, and the obliquity $\theta$ is equal
to $\cos^{-1}(\hat{n}_\star \bigcdot \hat{n}_{\rm o})$.
The vMF distribution is a widely-used model in directional statistics that
resembles a two-dimensional Gaussian distribution wrapped around a 
sphere.  Just as the Gaussian distribution
has the maximum entropy for a given variance,
the vMF distribution has the
maximum entropy for a fixed value of the mean obliquity \citep{Mardia1975}.
As $\kappa\rightarrow 0$, the distribution becomes isotropic,
and as $\kappa \rightarrow \infty$, it
approaches a delta-function centered on $\hat{n}_{\rm o}$.

We numerically computed the relationship between $\kappa$ and
the mean obliquity $\langle \theta\rangle$, as well as
$\langle \sin i\rangle$, assuming that $\Omega$ is uniformly
distributed between $0^\circ$ and $360^\circ$.
The results are shown in Figure~6, along with the
constraints on $\langle \theta\rangle$ obtained from our best-fitting models
of the data.
When all of the planet hosts are modeled together (premise 5a),
the 1-$\sigma$ allowed range for $\kappa$ is
from 1.7 to 4.2, and the mean obliquity ranges
from 37 to 58 degrees.
When the planet hosts are divided into two samples according
to effective temperature (premise 5b), the stars cooler than 6250\,K
have 1-$\sigma$ ranges of $\kappa=3.8$-16 and $\langle\theta\rangle = 18$-38 degrees,
while the ranges for the hotter stars 
are $\kappa = 0$-2.3 and $\langle\theta\rangle = 49$-88 degrees.

These results can be compared to previous inferences of $\kappa$
from different techniques and different samples of planet-hosting stars.
\cite{FabryckyWinn2009}
found $\kappa>7.6$ (95\% conf.) based on the first 11 observations
of the Rossiter-McLaughlin effect, all of which were hot Jupiter
hosts. Since that time, many more misaligned hot Jupiters
have been found; a more up-to-date analysis by \cite{MunozPerets2018}
gave $\kappa=2.2_{-0.6}^{+0.2}$.  This is comparable to the obliquity distribution
of the hotter half of the stars in our sample, while the cooler
half of the stars have a greater tendency to be well-aligned.

Previous inferences of the obliquity
distribution of {\it Kepler}
stars have mainly focused on the subset of stars with
detected rotation periods.  Such samples may be suffer
from biases related to orientation and transiting planet detection,
as noted in the Introduction. 
Nevertheless, in practice, our results are in agreement with the
prior results.  Since the stars in the previous
studies were almost all cooler than $6250$\,K,
the appropriate comparison
is to cooler half of our sample, for which we obtained
$\kappa=3.8$-16.
\cite{MortonWinn2014} analyzed
70 {\it Kepler} stars, finding $\kappa=19_{-12}^{+73}$
for stars with multiple transiting planets, and
$4.8_{-1.6}^{+2.0}$ for stars with only one detected
transiting planet.  To these results, \cite{Campante+2016} added
asteroseismic determinations of $\sin i$ for 25 {\it Kepler} stars,
finding $\kappa=11.5_{-5.7}^{+7.5}$ for the entire sample.
\cite{Winn+2017} expanded the work by \cite{MortonWinn2014} to include
156 stars and found $\kappa \gtrsim 5$ regardless of transit multiplicity. 
Likewise, \cite{MunozPerets2018} analyzed a sample
of 257 cool {\it Kepler} stars, and found $\kappa=14.5_{-6}^{+13.5}$.
All of the confidence intervals of these previous studies
overlap with ours, although in many cases the intervals are large.

\subsection{Model validation}
\label{subsec:validation}

\begin{figure}
\label{fig:SimulatedData}
\begin{center}
\includegraphics[width=0.45\textwidth]{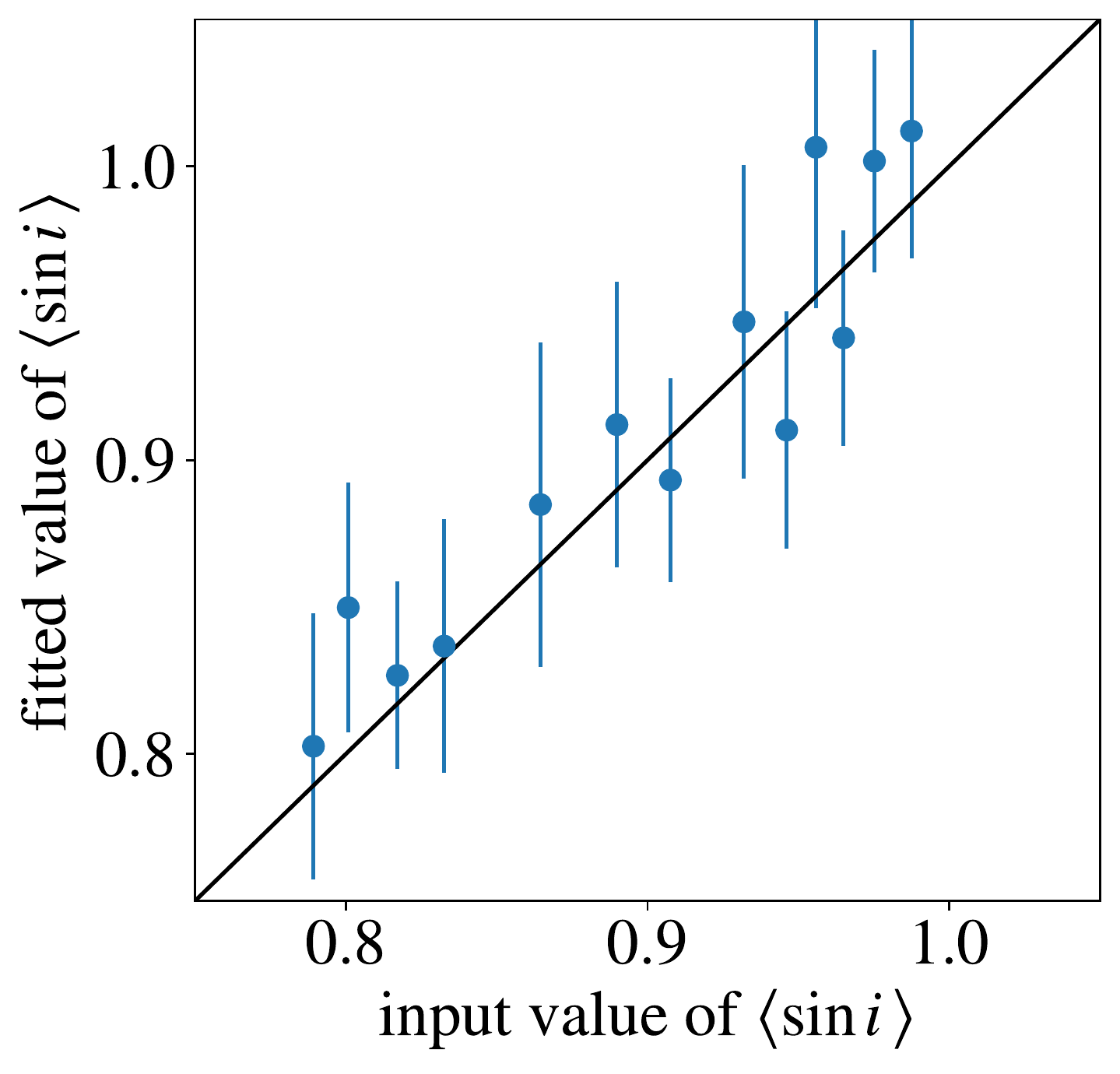}
\end{center}
\caption{Results of applying our modeling procedure to simulated
data with different assumed values of $\langle \sin i\rangle$,
as described in Section~\ref{subsec:validation}. The fitted
values and uncertainties are consistent with the input values
across the range of possible values.}
\end{figure}

To validate our modeling procedure, we fitted simulated data sets.
We generated simulated data sets with different input values
of $\langle \sin i\rangle$, and used our modeling procedure to ``recover'' the
best-fitting value of $\langle \sin i\rangle$ and test for
agreement.
Each simulated data set was created
as follows.  The 101 control stars were assigned random orientations,
and the 151 planet hosts
were assigned fictitious obliquities
drawn from a vMF distribution.  For all the stars, a fictitious
position angle
was drawn from a uniform distribution. 
We assumed a quadratic relationship between $\langle v\rangle$ and $T_{\rm eff}$
based on the best-fitting model to the real data.  Then, we assigned
a $v\sin i$ value to each star:
\begin{equation}
    v\sin i = \langle v\rangle (1 + 0.2 x_1) \sin i + 
    (1\,{\rm km\,s}^{-1}) x_2,
\end{equation}
where $x_1$ and $x_2$ are independent random
draws from a standard normal distribution $\mathcal{N}(0,1)$, to account for
the intrinsic dispersion of rotation velocities (assumed
to be 20\%) and the measurement
uncertainty, respectively.

Simulated data sets were created based on values of $\kappa$ ranging
from 0 to 40, corresponding to nearly the full range of possible
values of $\langle \sin i\rangle$.  We fitted the simulated data sets
using the same code that was used on the
actual data.  Figure~7 shows the results: the recovered values
of $\langle \sin i\rangle$ agree with the input values to within
the reported uncertainties, providing support for the validity of our
procedure.

\section{Discussion} \label{sec:end}

Overall, the {\it Kepler} planet-hosting
stars of spectral types from early-G to late-F have systematically higher
values of the projected rotation velocity than similar
stars chosen without regard to planets or spin-axis orientation.
Although this trend had been seen by \citet{Winn+2017}
and \citet{MunozPerets2018},
the improved control sample makes it possible
to be more confident in quantitative comparisons.
To explain the difference in terms of geometry,
the obliquity distribution of the planet hosts
must be intermediate between the limiting cases
of perfect alignment and random directions.

We analyzed the data using simple
models for the obliquity distribution, and presented evidence
that the hottest stars in the sample
have a broader distribution than the less hot stars.
Regardless of the details,
the important point is that many of the stars in our sample
(especially the late-F stars)
appear to have larger obliquities than the Sun.\footnote{The Sun's obliquity is
$6^\circ$ with respect to the total orbital angular momentum vector of the
8 planets, which is dominated by the contribution from Jupiter.}
This has been known for a decade for the hosts of hot Jupiters,
but to this point it has not been clear that it is also generally true of
the hosts of other types of planets.

A key assumption in our study is that the rotation velocities of
the planet hosts and control stars
are drawn from the same distribution.
We tried to ensure this is the case through careful matching
of observable spectroscopic parameters. Still, it remains
possible that systematic differences exist.
In principle, the control stars, being situated in a different
and more nearby location in the Galaxy, may have systematically
different rotation velocities than the planet hosts
even for fixed values of the spectroscopic parameters,
due to subtle differences in chemical composition or
formation history.
There might also be physical processes specific
to the formation and evolution of
{\it Kepler}-type planets that
alter a star's rotational history.
Tidal interactions with the known
planets are generally too weak to affect the star's rotation,
but one might speculate about previously ingested planets,
or differing magnetic and accretion histories.
Any such differences would be muted, though, by the fact
that between one-third and one-half of the control stars also have
{\it Kepler}-type planets that do not happen
to be transiting.

Despite these caveats,
our conclusions are supported by two
complementary lines of evidence.
The first is the work by \citet{Mazeh+2015}, noted
in Section 1.
They studied the obliquities of {\it Kepler} stars
using photometric variability data.
Stronger variability is expected for
stars viewed at high inclination,
the perspective that allows spots and plages to rotate into and out
of view.
Therefore, if the transiting-planet hosts have low obliquities,
they should show stronger variability than a sample of randomly-oriented stars,
whereas for random obliquities,
the planet hosts would show the same
level of variability as the randomly-oriented stars.
For stars cooler than about 6000\,K, 
\citet{Mazeh+2015} found the planet hosts to show stronger variability
than stars without detected transiting planets, by approximately the factor
of $4/\pi \approx 1.3$ that is expected if the planet
hosts have low obliquities
and the other stars are randomly oriented.

They also found that this trend reverses
for the hotter stars: the planet hosts
display {\it weaker} variability than the randomly-oriented stars,
with an amplitude ratio of 0.6.
This was surprising, because even in the seemingly extreme case
in which the planet hosts are randomly oriented, the amplitude
ratio would be 1.0.
For this ratio to fall below unity due only to differences
in viewing angles, we would be led to the unexpected conclusion that
the obliquities of the hot stars are preferentially near 90$^\circ$.

However, \citet{Mazeh+2015} found that at least part
of the difference between the variability levels
of the hot planet hosts and the randomly-oriented stars
is due to a selection effect.  Namely,
transiting planets are more readily detected around stars
with intrinsically lower levels of photometric variability.
Simulations of this selection effect showed that
it was indeed significant, but not large enough
to have reduced an intrinsic amplitude ratio of 1.3 all the way
down to the observed ratio of 0.6.
Thus, this study left
open the possibility that the hot {\it Kepler} stars have
a broader obliquity distribution than the cool stars,
an interpretation that harmonizes with our findings.

The second line of evidence for high obliquities among
hot stars with planets other than hot Jupiters comes
from recent observations of individual systems.
We are aware of only two obliquity measurements
for stars with effective temperatures between
5900\,K and 6450\,K that do not involve hot Jupiters, and
in both cases, the obliquity is high.
The first case is Kepler-408 ($T_{\rm eff} = 6088$\,K), which
has an Earth-sized planet in a 2.5-day orbit.
Asteroseismology revealed that the obliquity is
approximately 45$^\circ$ \citep{Kamiaka+2019}.
The second case is K2-290 ($T_{\rm eff} = 6302$\,K), which 
has a 3\,$R_\oplus$ planet in a 9.2-day orbit and a Jupiter-sized
planet in a 48-day orbit. Observations of the Rossiter-McLaughlin
effect show that the star's rotation is retrograde
(Hjorth et al., submitted).
Another relevant case is
Kepler-56 \citep{Huber+2013}, which has two planets of 
sizes 6.5 and 9.8~$R_\oplus$ and orbital periods of 10.5 and 21 days.
The host star is a subgiant with a mass of 1.3~$M_\odot$ and
an effective temperature of 4840\,K, although it was
probably about 6400\,K when it was on the main
sequence. The stellar obliquity is at least 45$^\circ$,
based on an asteroseismic analysis.
Some day we may accumulate enough 
of these individual measurements
to measure the obliquity distribution more directly.

While this is not the place to evaluate specific theories in detail,
we can list the previously published theories for obliquity excitation
that have the desired property that they
do not require the presence of a close-orbiting giant planet:\\[0.05in]
\indent $\bullet$ A misalignment between the protoplanetary disk
    due to inhomogeneities in the molecular cloud
    \citep{Bate2010,Fielding+2015, Takaishi2020},
    magnetic interactions \citep{Lai2011}, or a companion star \citep{Batygin2012,SpaldingBatygin2015,ZanazziLai2018}.\\[0.05in]
\indent $\bullet$ Ongoing nodal precession driven by a stellar
    companion or wide-orbiting giant planet on a highly inclined orbit \citep{AndersonLai2018}.\\[0.05in]
\indent $\bullet$ A resonance between the nodal precession rates of an inner
    planet and an outer planet that occurs during the dissipation
    of the protoplanetary disk \citep{Petrovich+2020}.\\[0.05in]
\indent $\bullet$ Random tumbling of the spin-axis orientation of the photosphere
    due to stochastic internal gravity waves \citep{RogersLin2012}.\\[0.05in]

Another desired property is that cooler stars with small planets should have low obliquities.  The dividing line of about $6250$\,K
is significant in stellar-evolutionary theory
because
the hot stars have thin or absent outer convective zones,
leading to weaker or absent magnetic braking,
more rapid rotation, and weaker tidal dissipation. Thus, it seems
likely that a successful theory will involve these
distinctions.
At least two of the theories listed above
make an explicit distinction between cool and hot stars:
those of \cite{RogersLin2012} (which pertains only to hot stars)
and \cite{SpaldingBatygin2015} (which appeals to the weaker magnetic
field of hot stars).
Of course, obliquities might be excited
and damped by different mechanisms in different situations,
including some that theoreticians have not yet identified.

\acknowledgements We are grateful to the anonymous referee for
a helpful critique of the manuscript, and to
Subo Dong for providing the LAMOST data in a convenient format. 
J.N.W.\ thanks the members of the Princeton exoplanet discussion group and Heather
Knutson's group for useful feedback,
and Geoff Marcy for input at the outset of this project.
J.N.W.\ also
acknowledges support from a NASA Keck PI Data Award, administered by
the NASA Exoplanet Science Institute.
S.A.\ acknowledges support from the Danish Council for Independent Research through the DFF Sapere Aude Starting Grant No.\ 4181-00487B,
and the Stellar Astrophysics Centre for which
funding is provided by The Danish National Research Foundation
(Grant agreement no.\ DNRF106).
M.R.K.\ is supported by the NSF Graduate Research Fellowship grant no.\ DGE\,1339067. 
This work made use of data from the LAMOST (Guoshoujing) Telescope, a National Major Scientific Project built by the Chinese Academy of Sciences, for which funding was provided by the National Development and Reform Commission. LAMOST is operated and managed by the National Astronomical Observatories, Chinese Academy of Sciences.
We acknowledge the very significant cultural role and reverence that the summit of Maunakea has always had within the indigenous Hawaiian community. We are most fortunate to have the opportunity to conduct observations from this mountain.

\facility{Keck:I (HIRES)}
\facility{LAMOST}

%\clearpage

\startlongtable
\begin{deluxetable}{cccccc}
\label{tbl:planet_hosts}
\tablecaption{Spectroscopic properties of the planet hosts.}
\tablehead{
  KIC no. & KOI no. & $T_\mathrm{eff}$ [K] & $\log g$ & [Fe/H] & $v\sin i$ [km~s$^{-1}$]
}
\startdata
\input{planet_hosts}
\enddata
\tablecomments{The uncertainties in $T_\mathrm{eff}$, $\log g$, [Fe/H], and $v\sin i$ are 60\,K, 0.1, 0.1, and 1\,km~s$^{-1}$, respectively.}
\end{deluxetable}

\startlongtable
\begin{deluxetable}{ccccc}
\label{tbl:control_stars}
\tablecaption{Spectroscopic properties of the control stars.}
\tablehead{
  KIC no. & $T_\mathrm{eff}$ [K] & $\log g$ & [Fe/H] & $v\sin i$ [km~s$^{-1}$]
}
\startdata
\input{control_stars.tex}
\enddata
\tablecomments{The uncertainties in $T_\mathrm{eff}$, $\log g$, [Fe/H], and $v\sin i$ are 60\,K, 0.1, 0.1, and 1\,km/s, respectively.}
\end{deluxetable}

\bibliography{papers}{}
\bibliographystyle{aasjournal}

\end{document}

%% file: planet_hosts.tex
  1724719 &       4212 &     6215 &     4.35 &   $0.14$ &       6.56 \\
  1871056 &       1001 &     6298 &     4.15 &   $0.14$ &       8.97 \\
  1996180 &       2534 &     6118 &     4.36 &   $0.02$ &       4.43 \\
  2142522 &       2403 &     6186 &     4.37 &   $0.18$ &       5.34 \\
  2307415 &       2053 &     6183 &     4.27 &   $0.08$ &       6.77 \\
  2854914 &       1113 &     6099 &     4.27 &   $0.04$ &       5.24 \\
  2989404 &       1824 &     5987 &     4.35 &  $-0.16$ &       3.07 \\
  3114811 &       1117 &     6401 &     4.12 &  $-0.05$ &       9.61 \\
  3120904 &       3277 &     5997 &     3.95 &  $-0.02$ &       7.33 \\
  3447722 &       1198 &     6323 &     4.39 &   $0.10$ &       8.36 \\
  3632418 &        975 &     6186 &     4.08 &  $-0.05$ &       7.31 \\
  3642289 &        301 &     5995 &     4.10 &  $-0.08$ &       5.51 \\
  3656121 &        386 &     5983 &     4.40 &   $0.06$ &       3.60 \\
  3661886 &       2279 &     5950 &     4.25 &   $0.05$ &       2.60 \\
  3867615 &       2289 &     6048 &     4.28 &   $0.09$ &       4.90 \\
  3939150 &       1215 &     6054 &     4.02 &   $0.09$ &       2.90 \\
  3964109 &        393 &     6280 &     4.34 &  $-0.19$ &       7.99 \\
  3969687 &       2904 &     6077 &     3.96 &   $0.30$ &       7.47 \\
  4242692 &       3928 &     6192 &     4.31 &  $-0.05$ &       6.59 \\
  4278221 &       1615 &     6013 &     4.43 &   $0.22$ &       7.88 \\
  4349452 &        244 &     6280 &     4.31 &  $-0.03$ &       9.25 \\
  4478168 &        626 &     6131 &     4.33 &  $-0.17$ &       3.99 \\
  4563268 &        627 &     6050 &     4.18 &   $0.18$ &       5.39 \\
  4656049 &        629 &     6059 &     4.34 &  $-0.29$ &       2.65 \\
  4741126 &       1534 &     6208 &     4.29 &   $0.04$ &       8.07 \\
  4833421 &        232 &     5994 &     4.25 &  $-0.06$ &       4.03 \\
  4914566 &       2635 &     6052 &     4.09 &  $-0.14$ &       6.99 \\
  5120087 &        639 &     6184 &     4.34 &   $0.05$ &       7.39 \\
  5175024 &       2563 &     6119 &     4.00 &   $0.11$ &       7.53 \\
  5183357 &       1669 &     6140 &     4.24 &  $-0.05$ &       6.94 \\
  5281113 &       4411 &     6119 &     4.44 &  $-0.14$ &       4.18 \\
  5350244 &       2555 &     6180 &     4.38 &  $-0.07$ &       6.14 \\
  5384079 &       2011 &     6455 &     4.44 &  $-0.02$ &      15.21 \\
  5514383 &        257 &     6220 &     4.36 &   $0.12$ &       7.49 \\
  5561278 &       1621 &     6089 &     4.06 &   $0.02$ &       5.90 \\
  5613330 &        649 &     6244 &     4.10 &   $0.22$ &       7.80 \\
  5631630 &       2010 &     6416 &     4.27 &  $-0.00$ &      10.03 \\
  5866724 &         85 &     6229 &     4.22 &   $0.15$ &       8.71 \\
  5880320 &       1060 &     6351 &     4.16 &  $-0.29$ &      12.17 \\
  5966154 &        655 &     6171 &     4.25 &  $-0.02$ &       2.69 \\
  6026924 &       4276 &     6221 &     4.00 &   $0.12$ &       8.20 \\
  6105462 &       2098 &     6305 &     4.33 &  $-0.13$ &      11.39 \\
  6125481 &        659 &     6374 &     4.45 &  $-0.03$ &       8.77 \\
  6196457 &        285 &     5952 &     4.03 &   $0.20$ &       3.96 \\
  6206214 &       2252 &     6116 &     4.02 &   $0.08$ &       7.61 \\
  6269070 &       2608 &     6429 &     4.40 &   $0.06$ &      18.06 \\
  6289257 &        307 &     6035 &     4.41 &  $-0.08$ &       5.44 \\
  6310636 &       1688 &     5993 &     4.08 &   $0.18$ &       5.64 \\
  6345732 &       2857 &     6189 &     4.15 &   $0.24$ &       6.28 \\
  6442377 &        176 &     6423 &     4.44 &  $-0.19$ &      11.03 \\
  6599975 &       3438 &     6054 &     4.38 &  $-0.14$ &       3.72 \\
  6716545 &       2906 &     6087 &     4.15 &  $-0.28$ &       5.98 \\
  6937529 &       4382 &     6217 &     4.24 &  $-0.12$ &       9.98 \\
  7040629 &        671 &     6093 &     4.24 &   $0.03$ &       5.06 \\
  7133294 &       4473 &     5952 &     4.00 &   $0.04$ &       5.83 \\
  7175184 &        369 &     6227 &     4.37 &  $-0.12$ &       6.86 \\
  7215603 &       1618 &     6209 &     4.16 &   $0.13$ &      10.07 \\
  7219825 &        238 &     6131 &     4.27 &  $-0.10$ &       5.88 \\
  7259298 &       2561 &     6014 &     4.16 &   $0.04$ &       4.12 \\
  7375348 &        266 &     6232 &     4.42 &  $-0.02$ &       5.53 \\
  7673192 &       2722 &     6140 &     4.38 &  $-0.07$ &       6.26 \\
  7755636 &       1921 &     6479 &     4.33 &   $0.04$ &      18.56 \\
  7831264 &        171 &     6237 &     4.30 &   $0.10$ &       7.88 \\
  8013439 &       2352 &     6387 &     4.29 &  $-0.14$ &       7.17 \\
  8073705 &       3245 &     6115 &     4.31 &  $-0.10$ &       3.57 \\
  8077137 &        274 &     6081 &     4.09 &  $-0.04$ &       6.07 \\
  8081187 &       1951 &     6093 &     4.33 &  $-0.20$ &       4.35 \\
  8121310 &        317 &     6520 &     4.28 &   $0.05$ &      15.94 \\
  8158127 &       1015 &     5950 &     4.16 &  $-0.10$ &       4.37 \\
  8161561 &        688 &     6218 &     4.27 &  $-0.06$ &       8.88 \\
  8193178 &        572 &     6003 &     4.16 &  $-0.06$ &       4.28 \\
  8212002 &       2593 &     6238 &     4.28 &   $0.22$ &       7.46 \\
  8292840 &        260 &     6292 &     4.34 &  $-0.16$ &       8.71 \\
  8394721 &        152 &     6428 &     4.39 &   $0.07$ &      14.05 \\
  8410727 &       1148 &     6127 &     4.27 &   $0.16$ &       5.51 \\
  8494142 &        370 &     6117 &     4.09 &   $0.07$ &       6.71 \\
  8636434 &       3946 &     6325 &     4.26 &  $-0.22$ &       7.13 \\
  8644365 &       3384 &     5956 &     4.18 &   $0.05$ &       3.94 \\
  8738735 &        693 &     6018 &     4.08 &  $-0.03$ &       6.54 \\
  8751796 &       3125 &     6293 &     4.32 &  $-0.11$ &       9.55 \\
  8773015 &       4301 &     6266 &     4.44 &  $-0.14$ &       5.04 \\
  8822366 &       1282 &     6087 &     4.16 &  $-0.13$ &       6.22 \\
  8883329 &       2595 &     6332 &     4.30 &   $0.04$ &      14.37 \\
  8972058 &        159 &     6055 &     4.40 &  $-0.01$ &       3.46 \\
  9009036 &       4585 &     6158 &     4.18 &  $-0.11$ &       9.50 \\
  9015738 &       1616 &     6067 &     4.29 &   $0.18$ &       5.32 \\
  9026749 &       2564 &     6087 &     4.03 &  $-0.13$ &       9.13 \\
  9070666 &       3008 &     5981 &     4.07 &  $-0.08$ &       5.44 \\
  9277896 &       1632 &     6088 &     4.16 &   $0.02$ &       7.60 \\
  9412623 &       4640 &     6396 &     4.28 &  $-0.06$ &      10.71 \\
  9450647 &        110 &     6241 &     4.38 &  $-0.16$ &       6.95 \\
  9451706 &        271 &     6158 &     4.23 &   $0.27$ &       6.56 \\
  9458613 &        707 &     5953 &     4.07 &   $0.14$ &       4.36 \\
  9466429 &       2786 &     6367 &     4.25 &  $-0.13$ &       7.52 \\
  9467404 &       2717 &     6025 &     4.25 &  $-0.02$ &       4.61 \\
  9529744 &       1806 &     6146 &     4.27 &  $-0.05$ &       6.94 \\
  9530945 &        708 &     6100 &     4.19 &  $-0.10$ &       7.11 \\
  9549648 &       1886 &     6221 &     4.18 &   $0.05$ &      10.52 \\
  9579641 &        115 &     5961 &     4.39 &  $-0.08$ &       2.94 \\
  9590976 &        710 &     6540 &     4.26 &  $-0.06$ &      14.96 \\
  9649706 &       2049 &     5972 &     4.15 &   $0.07$ &       4.14 \\
  9696358 &       2545 &     6105 &     3.99 &   $0.14$ &       6.86 \\
  9717943 &       2273 &     6038 &     4.25 &   $0.25$ &       4.55 \\
  9763348 &       1852 &     6415 &     4.35 &  $-0.17$ &       4.18 \\
  9782691 &        590 &     5981 &     4.42 &  $-0.00$ &       3.72 \\
  9881662 &        327 &     6043 &     4.28 &  $-0.02$ &       4.35 \\
  9886361 &       2732 &     6082 &     4.07 &   $0.08$ &       4.90 \\
  9892816 &       1955 &     6249 &     4.25 &   $0.09$ &       9.52 \\
  9904006 &       2135 &     6171 &     4.20 &   $0.05$ &       7.60 \\
  9965439 &        722 &     6188 &     4.38 &  $-0.11$ &       5.54 \\
 10227020 &        730 &     5952 &     4.19 &   $0.27$ &       2.45 \\
 10253547 &       2153 &     6023 &     4.19 &  $-0.06$ &       5.79 \\
 10337258 &        333 &     6237 &     4.41 &   $0.11$ &       8.23 \\
 10460984 &        474 &     5981 &     4.42 &   $0.02$ &       3.28 \\
 10471515 &       2961 &     6078 &     4.27 &   $0.13$ &       5.24 \\
 10615440 &       4765 &     6227 &     4.31 &   $0.15$ &       6.76 \\
 10916600 &       2623 &     6178 &     4.10 &   $0.13$ &       7.18 \\
 10963065 &       1612 &     6095 &     4.27 &  $-0.14$ &       2.83 \\
 11019987 &       3060 &     6187 &     4.10 &  $-0.06$ &       6.17 \\
 11043167 &       1444 &     6191 &     4.11 &   $0.10$ &       8.35 \\
 11086270 &        124 &     5977 &     4.27 &  $-0.10$ &       3.33 \\
 11121752 &       2333 &     6081 &     4.35 &  $-0.15$ &       3.66 \\
 11127479 &       2792 &     5969 &     4.17 &   $0.21$ &       4.26 \\
 11133306 &        276 &     5993 &     4.34 &  $-0.01$ &       2.43 \\
 11259686 &        294 &     6076 &     4.36 &   $0.05$ &       4.39 \\
 11336883 &       1445 &     6351 &     4.26 &   $0.14$ &      13.81 \\
 11337566 &       2632 &     6209 &     4.07 &  $-0.02$ &      10.31 \\
 11342416 &       2366 &     6165 &     4.31 &   $0.01$ &       6.64 \\
 11401755 &        277 &     5987 &     4.13 &  $-0.19$ &       4.84 \\
 11401767 &       2195 &     6130 &     4.24 &  $-0.12$ &       4.96 \\
 11442793 &        351 &     5993 &     4.24 &   $0.09$ &       3.64 \\
 11457726 &       2047 &     6172 &     4.37 &  $-0.02$ &       5.33 \\
 11460462 &       2110 &     6404 &     4.28 &   $0.23$ &      16.45 \\
 11499228 &       2109 &     6028 &     4.07 &   $0.04$ &       7.41 \\
 11560897 &       2365 &     5952 &     4.17 &   $0.21$ &       3.66 \\
 11572193 &       3109 &     6237 &     4.05 &   $0.09$ &      17.29 \\
 11621223 &        355 &     6122 &     4.25 &   $0.20$ &       4.94 \\
 11656246 &       1532 &     6143 &     4.21 &   $0.15$ &       6.59 \\
 11666881 &        167 &     6209 &     4.15 &   $0.04$ &       2.43 \\
 11807274 &        262 &     6171 &     4.18 &  $-0.09$ &      10.01 \\
 11811193 &       2260 &     6188 &     4.22 &   $0.01$ &       7.88 \\
 11905011 &        297 &     6181 &     4.31 &   $0.12$ &       6.96 \\
 12024120 &        265 &     6015 &     4.18 &   $0.14$ &       2.31 \\
 12058931 &        546 &     5971 &     4.24 &   $0.12$ &       3.55 \\
 12120484 &       2407 &     5956 &     4.12 &   $0.19$ &       5.19 \\
 12206313 &       2714 &     5989 &     4.02 &  $-0.02$ &       5.79 \\
 12254909 &       2372 &     5970 &     4.17 &  $-0.09$ &       3.92 \\
 12314973 &        279 &     6294 &     4.27 &   $0.18$ &      12.93 \\
 12416661 &       3122 &     6177 &     4.01 &   $0.23$ &      10.13 \\
 12600735 &        548 &     6000 &     4.23 &  $-0.06$ &       3.04

%% file: control_stars.tex
  2158850 &     6039 &     4.35 &  $-0.10$ &       2.68 \\
  2998253 &     6292 &     4.32 &   $0.11$ &       9.15 \\
  3123191 &     6313 &     4.40 &  $-0.06$ &       8.00 \\
  3338777 &     5994 &     4.06 &   $0.04$ &       6.93 \\
  3831297 &     6112 &     4.37 &  $-0.08$ &       3.65 \\
  3936993 &     6176 &     4.12 &  $-0.12$ &       7.90 \\
  4346201 &     6095 &     4.14 &  $-0.21$ &       5.14 \\
  4484238 &     6170 &     4.26 &   $0.07$ &       7.31 \\
  4645245 &     6119 &     4.41 &  $-0.16$ &       4.97 \\
  4753390 &     6379 &     4.27 &   $0.16$ &      11.03 \\
  5094944 &     6442 &     4.39 &   $0.12$ &      13.67 \\
  5183581 &     6325 &     4.14 &  $-0.04$ &      10.82 \\
  5184384 &     6072 &     4.19 &  $-0.00$ &       5.97 \\
  5468089 &     6209 &     4.11 &   $0.04$ &       8.91 \\
  5510904 &     6431 &     4.44 &   $0.11$ &      14.67 \\
  5788360 &     6195 &     4.27 &   $0.10$ &       8.02 \\
  5803208 &     6440 &     4.28 &   $0.26$ &      14.11 \\
  5856836 &     6287 &     4.28 &   $0.17$ &       7.64 \\
  5865892 &     6407 &     4.44 &  $-0.05$ &       7.75 \\
  6314137 &     6390 &     4.40 &  $-0.05$ &       9.46 \\
  6364123 &     6114 &     4.16 &  $-0.14$ &       4.00 \\
  6425358 &     6167 &     4.26 &   $0.16$ &       4.82 \\
  6438107 &     6077 &     4.05 &   $0.02$ &       5.54 \\
  6680045 &     6161 &     4.32 &  $-0.03$ &       4.98 \\
  6689943 &     6058 &     4.32 &   $0.04$ &       5.89 \\
  6778540 &     6031 &     4.35 &   $0.04$ &       3.60 \\
  7094508 &     6245 &     4.03 &   $0.19$ &       6.97 \\
  7206837 &     6430 &     4.28 &   $0.22$ &       8.75 \\
  7260381 &     6113 &     4.35 &   $0.07$ &       4.25 \\
  7383120 &     6089 &     4.18 &  $-0.11$ &       3.54 \\
  7422905 &     5993 &     4.41 &  $-0.15$ &       1.71 \\
  7434909 &     5976 &     4.27 &   $0.17$ &       2.29 \\
  7465902 &     6172 &     4.25 &   $0.17$ &       6.53 \\
  7670943 &     6372 &     4.37 &  $-0.04$ &      11.84 \\
  7811344 &     6263 &     4.24 &   $0.08$ &       9.11 \\
  7880676 &     6180 &     4.06 &   $0.14$ &       5.30 \\
  8013078 &     6091 &     4.13 &  $-0.15$ &       3.18 \\
  8017790 &     6042 &     4.35 &   $0.03$ &       4.07 \\
  8077525 &     6237 &     4.32 &   $0.01$ &       3.51 \\
  8112746 &     6126 &     4.24 &   $0.06$ &       3.48 \\
  8228742 &     6068 &     4.01 &  $-0.06$ &       4.25 \\
  8289241 &     6249 &     4.29 &   $0.10$ &       6.96 \\
  8420801 &     6226 &     4.13 &   $0.07$ &       7.24 \\
  8493800 &     6034 &     4.18 &   $0.05$ &       5.72 \\
  8494872 &     6079 &     4.35 &  $-0.13$ &       4.88 \\
  8623058 &     6243 &     4.24 &   $0.19$ &       5.51 \\
  8650186 &     6172 &     4.22 &   $0.04$ &       4.93 \\
  8696343 &     5989 &     4.24 &   $0.07$ &       4.45 \\
  8717023 &     6169 &     4.30 &   $0.21$ &       4.94 \\
  8973900 &     6345 &     4.19 &   $0.20$ &      10.75 \\
  9007356 &     5961 &     4.13 &   $0.09$ &       3.64 \\
  9157245 &     5970 &     4.18 &   $0.09$ &       4.10 \\
  9225600 &     6245 &     4.16 &   $0.27$ &       6.24 \\
  9273544 &     6053 &     4.21 &  $-0.16$ &       3.64 \\
  9289275 &     6086 &     4.08 &   $0.17$ &       6.95 \\
  9329766 &     5966 &     4.42 &  $-0.12$ &       3.05 \\
  9347707 &     6491 &     4.33 &   $0.23$ &      18.22 \\
  9390670 &     6382 &     4.24 &   $0.24$ &      15.92 \\
  9402649 &     6130 &     4.25 &  $-0.13$ &       5.53 \\
  9468847 &     5964 &     3.99 &  $-0.06$ &       6.59 \\
  9529969 &     5953 &     4.16 &  $-0.25$ &       2.96 \\
  9579208 &     6246 &     4.20 &  $-0.19$ &       7.80 \\
  9592705 &     6197 &     4.04 &   $0.22$ &       7.91 \\
  9613220 &     6092 &     4.27 &  $-0.23$ &       2.94 \\
  9644337 &     6295 &     4.29 &   $0.10$ &       8.85 \\
  9651253 &     6130 &     4.38 &   $0.05$ &       4.78 \\
  9664404 &     5951 &     4.23 &   $0.08$ &       3.49 \\
  9754284 &     6048 &     4.07 &   $0.10$ &       3.95 \\
  9814780 &     6074 &     4.36 &  $-0.01$ &       3.38 \\
  9898249 &     6430 &     4.44 &  $-0.13$ &      13.71 \\
  9912680 &     6067 &     4.04 &   $0.06$ &       4.97 \\
 10024648 &     6252 &     4.11 &  $-0.01$ &       9.17 \\
 10025841 &     5996 &     4.07 &   $0.26$ &       4.04 \\
 10079226 &     5969 &     4.28 &   $0.17$ &       3.54 \\
 10097040 &     6246 &     4.27 &   $0.25$ &       6.24 \\
 10322381 &     6149 &     4.32 &  $-0.29$ &       4.47 \\
 10651962 &     6186 &     4.26 &  $-0.04$ &       5.12 \\
 10813492 &     6072 &     4.12 &   $0.09$ &       6.20 \\
 10813660 &     6138 &     4.21 &  $-0.03$ &       5.35 \\
 10907438 &     6122 &     3.96 &   $0.18$ &       5.05 \\
 10988876 &     6133 &     4.33 &  $-0.10$ &       2.56 \\
 11025641 &     6110 &     4.19 &  $-0.17$ &       3.98 \\
 11136719 &     6536 &     4.25 &   $0.24$ &      16.31 \\
 11197632 &     6084 &     4.34 &   $0.05$ &       3.67 \\
 11389437 &     6060 &     4.10 &   $0.08$ &       5.66 \\
 11444160 &     6150 &     4.40 &   $0.04$ &       6.39 \\
 11467550 &     6112 &     4.04 &   $0.06$ &       7.12 \\
 11599875 &     6121 &     4.14 &   $0.09$ &       9.33 \\
 11654380 &     5972 &     4.10 &   $0.02$ &       5.40 \\
 11710285 &     6351 &     4.30 &  $-0.23$ &       6.92 \\
 11719930 &     6208 &     4.27 &  $-0.06$ &       6.73 \\
 11754352 &     6207 &     4.22 &   $0.26$ &       4.86 \\
 11763874 &     6024 &     4.20 &  $-0.04$ &       4.73 \\
 11870319 &     6125 &     4.31 &  $-0.11$ &       5.06 \\
 12006631 &     6144 &     4.41 &   $0.10$ &       3.07 \\
 12009504 &     6118 &     4.18 &  $-0.05$ &       6.70 \\
 12058453 &     6011 &     4.17 &   $0.05$ &       4.04 \\
 12069127 &     6293 &     4.04 &   $0.14$ &       3.77 \\
 12117868 &     5968 &     4.20 &  $-0.03$ &       3.58 \\
 12555240 &     6120 &     4.30 &  $-0.14$ &       2.25 \\
 12785394 &     6520 &     4.42 &   $0.04$ &      17.72